\documentclass{article}
\usepackage{spconf,amsmath,graphicx}
\usepackage{cite}

\usepackage{xcolor}
\PassOptionsToPackage{hyphens}{url}\usepackage{hyperref}
\usepackage[T1]{fontenc}
\usepackage[latin9]{inputenc}
\usepackage{collcell}
\usepackage{hhline}
\usepackage{pgf}
\usepackage{multirow}
\newcolumntype{P}[1]{>{\centering\arraybackslash}p{#1}}

\title{Detecting COVID-19 and Community Acquired Pneumonia using \\Chest CT Scan Images with Deep Learning}
\name{
\begin{tabular}{@{}c@{}}
Shubham Chaudhary$^{\diamondsuit}$, Sadbhawna$^{\diamondsuit}$, Vinit Jakhetiya$^{\diamondsuit}$, Badri N Subudhi$^{\diamondsuit}$, \\ Ujjwal Baid$^{\dagger}$, Sharath Chandra Guntuku$^{\dagger}$
\end{tabular}}
  \address{$^{\diamondsuit}$ Indian Institute of Technology, Jammu, India \\
      $^{\dagger}$ University of Pennsylvania, PA, USA}
\usepackage{subfig}

\begin{document}
%
\maketitle
\begin{abstract}
We propose a two-stage Convolutional Neural Network (CNN) based classification framework for detecting COVID-19 and Community Acquired Pneumonia (CAP) using the chest Computed Tomography (CT) scan images. 
In the first stage, an infection - COVID-19 or CAP, is detected using a pre-trained DenseNet architecture. Then, in the second stage, a fine-grained three-way classification is done 
using EfficientNet architecture. The proposed COVID+CAP-CNN framework achieved a slice-level classification accuracy of over 94\% at identifying COVID-19 and CAP.
Further, the proposed framework has the potential to be an initial screening tool for differential diagnosis of COVID-19 and CAP, achieving a validation accuracy of over 89.3\% at the finer three-way COVID-19, CAP, and healthy classification. Within the IEEE ICASSP 2021 Signal Processing Grand Challenge (SPGC) on COVID-19 Diagnosis, our proposed two-stage classification framework achieved an overall accuracy of 90\% and sensitivity of .857, .9, and .942 at distinguishing COVID-19, CAP, and normal individuals respectively, to rank first in the evaluation. Code and model weights are available at ~\url{https://github.com/shubhamchaudhary2015/ct_covid19_cap_cnn}
\end{abstract}
\section{Introduction}
\label{sec:intro}

As of March 2021, there have been more than 119 million confirmed cases of the Severe Acute Respiratory Syndrome Coronavirus 2 (SARS-CoV-2) infection, the virus that causes the novel coronavirus disease (COVID-19), resulting in over 2.6 million reported deaths~\cite{dong2020interactive}. Chest Computed Tomography (CT) images have shown to be an essential method for detecting interstitial pneumonia, a distinctive feature of COVID-19~\cite{grillet2020acute}. Deep learning based computational imaging techniques are promising at the evaluation of positive COVID-19 cases~\cite{zheng2020deep}.

\begin{figure}[t!]
	\centering
	\captionsetup[subfloat]{oneside,margin={0.2cm,0.2cm}}
	\subfloat[Normal]{
		\includegraphics[height=28mm]{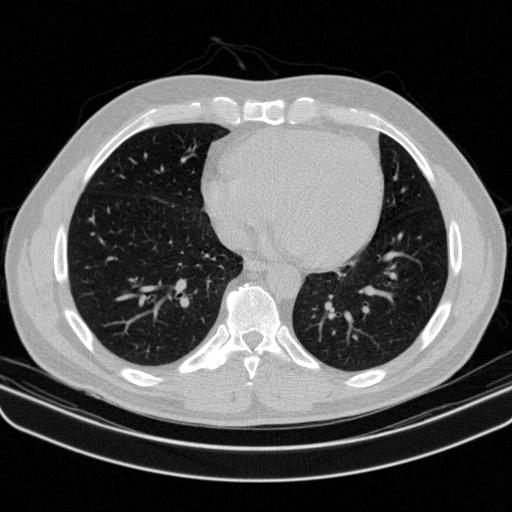} }
	\subfloat[COVID-19]{
		\includegraphics[height=28mm]{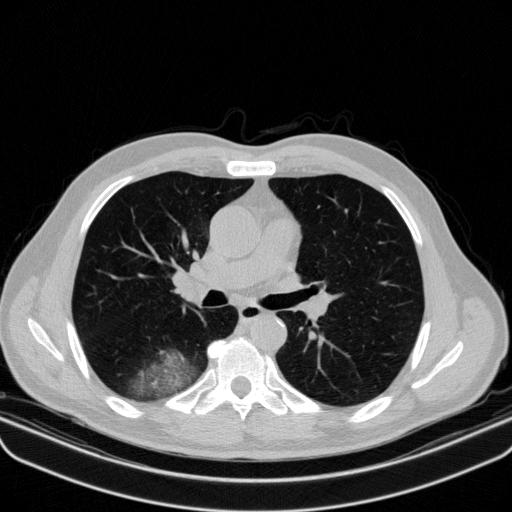} }
	\subfloat[CAP]{
		\includegraphics[height=28mm]{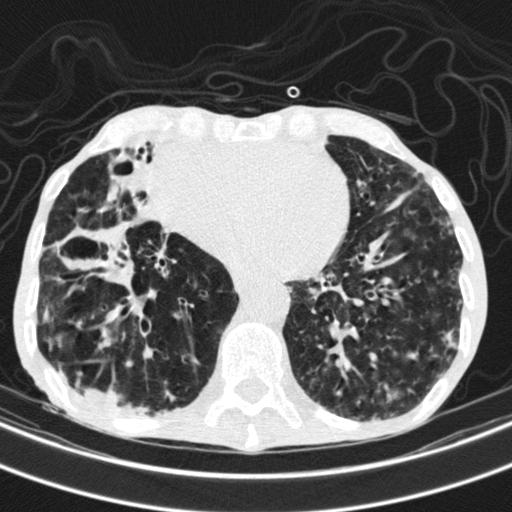}
	}
	\caption{Segmented slice images from  three example CT volumes of a) normal, b) COVID-19, c) CAP patients}
\end{figure}

Several works have studied the feasibility of automating COVID-19 detection using Convolutional Neural Networks (CNN)~\cite{covid-cd-md, nvidia_nature, waterloo_xray, uc_san_diago, wuhan, covid-caps, covid-fact, mixcaps}. Harmon \MakeLowercase{\textit{et al.}} showed that AI-based algorithms trained on a multinational cohort could classify CT images between pneumonia associated with COVID-19 and non-COVID-19 pneumonia with over 90\% accuracy~\cite{nvidia_nature}. 
Wang \MakeLowercase{\textit{et al.}} proposed a Deep tailored CNN for chest X-ray images to detect COVID-19~\cite{waterloo_xray} and released an open-source dataset of 13975 chest X-ray images. On another dataset of 349 COVID-19 CT images from 216 patients and 463 non-COVID-19 samples, an automatic diagnosis system using multi-task and self-supervised learning techniques reported accuracy 89\%~\cite{uc_san_diago}. Zheng \MakeLowercase{\textit{et al.}} used 499 3D CT volumes to predict COVID-19 infections by training a weakly-supervised deep learning model~\cite{wuhan}. COVID-CAPS~\cite{covid-caps} is a capsules network-based framework trained on X-ray images to identify COVID-19 infections. COVID-FACT~\cite{covid-fact} is a two-stage network where the first stage identifies slices with infection and classifies them further into COVID-19 and other infections in the second stage. 

Manual analysis of chest CT scans by professional medical professionals is a resource and time-intensive process. An automated risk assessment system could potentially serve as an initial screening tool to aid medical professionals' diagnosis. One of the challenges associated with the automatic detection of COVID-19 is its differential diagnosis compared to CAP~\cite{shi2021large}. To address this, we propose a two-stage Convolutional Neural Network (CNN) based framework, termed COVID+CAP-CNN, for the differential detection of COVID-19 and CAP.

\begin{figure*}[t!]
\centering
\subfloat{
  \includegraphics[width=\textwidth]{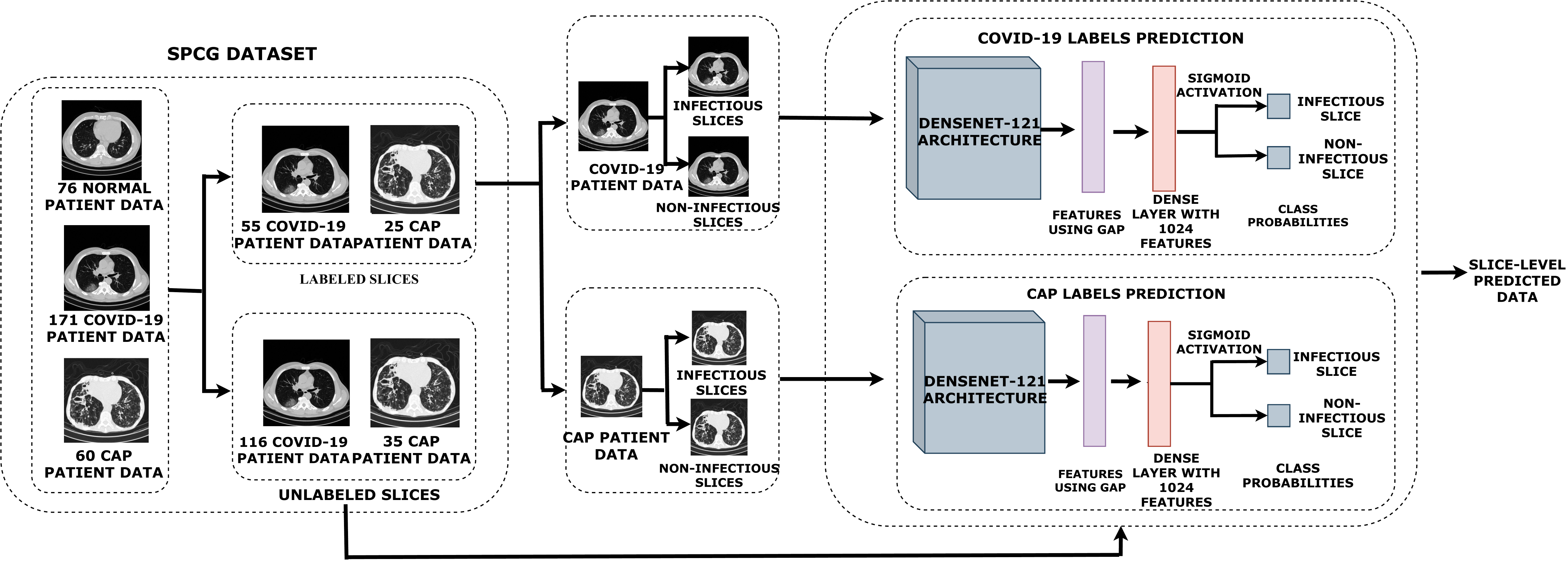}
}
\caption{Overview of Stage-1 of the proposed COVID+CAP-CNN Architecture}
\end{figure*}
In Stage-1, we train two different CNN architectures for labeling COVID-19 and CAP CT scans. In Stage-2, we propose to use EfficientNet architecture to classify individuals and CT scans into three classes, i.e., Normal, COVID-19, and CAP. We use the SPGC-COVID dataset~\cite{covid-cd-md}, which contains CT scans of individuals with a combination of COVID-19 and CAP, along with CT scans of individuals with no infections. The dataset has slice-level as well as patient-level labels for several individuals. In Fig. 1, example CT scans for normal individuals and patients affected with COVID-19 and CAP in the SPGC-COVID dataset are shown. Sample descriptives of the SPGC dataset are given in Table 1.

\begin{table}[h!]
\caption{ The description of SPGC dataset.}
\label{table}
\centering
\small
\setlength{\tabcolsep}{3pt}
\begin{tabular}{|c||P{2.5cm}|P{3.5cm}|}
\hline
\textbf{Disease Type}& \textbf{Number of Individuals} & \textbf{Number of Patients with slice-level labels} \\
\hline
Normal & 76 & -- \\
CAP & 60 & 25 \\
COVID-19 & 171 & 55 \\
\hline
\end{tabular}
\label{tab1}
\end{table}

\section{PROPOSED MODEL}

\begin{figure*}[t!]
\centering
\subfloat{
  \includegraphics[width=\textwidth]{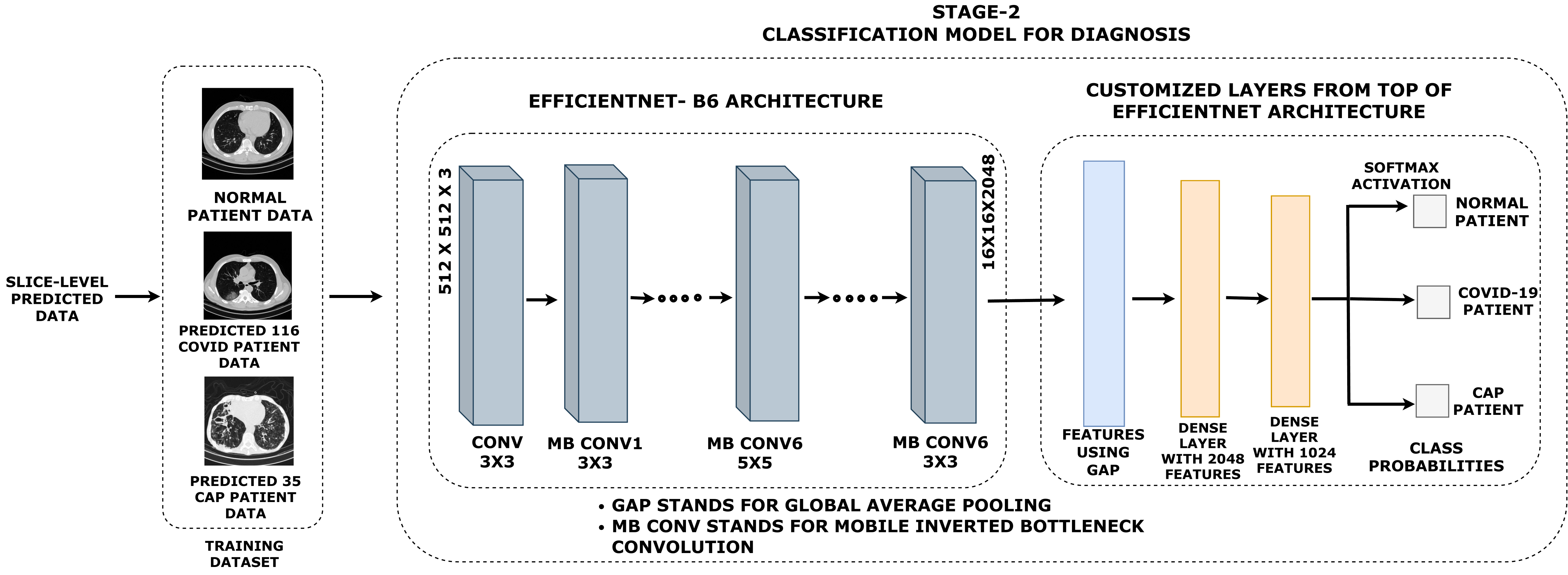}
}
\caption{Overview of Stage-2 of the proposed COVID+CAP-CNN Architecture.}
\end{figure*}

Our proposed COVID-19 and CAP detection system using deep learning, termed as COVID+CAP-CNN, consists of two stages. In the first stage, unlabelled CT scans in the SPGC dataset are labeled using a pre-trained CNN-based algorithm (detailed in subsection 2.2). In the second stage, labeled slices are pooled into individual-level estimates of COVID-19, CAP infected patients, and normal individuals using the state-of-the-art EfficientNet architecture (detailed in subsection 2.3). The complete architecture of the two stages of the proposed COVID-CNN model is shown in Figs. 2 and 3. We describe data pre-processing and implementation of these two stages in the following subsections.

\subsection{Dataset Pre-processing}
Each patient's data in a chest CT scan is given in the form of a 3D volume, a combination of many image slices captured from different angles. Images provided in the SPGC-COVID dataset were mapped from Hounsfield Units (HU) to [0, 255] using a window centered at -500 HU with a width of 1300 HU. Further, we observed that the crucial signal associated with infection is often present in the 3D chest CT volumes' central slices. Consequently, we selected the middle slices from the 3D volumes to fine-tune the proposed COVID-CNN architecture. For COVID-19 and CAP patients with over 80 slices in the 3D chest CT volume, we select the middle 80 slices, and if the number of slices is less than 80, then we take the middle 40 slices. The distribution of adopted data for Stage-1 architectures is shown in Table 2. To overcome class imbalance, a similar number of CT scans are selected from different classes. 

\begin{table}[h]
\caption{Data used for fine-tuning Stage-1}
\label{table}
\centering
\small
\setlength{\tabcolsep}{2pt}
\begin{tabular}{|c||c|c|}
\hline
\textbf{Disease Type}& \textbf{Number of Patients} &  \begin{tabular}{c}
                   \textbf{Total Selected Slices}\\
                   \hline
                   \begin{tabular}{P{1.3cm}|P{2cm}}
                   Infectious & Non-Infectious \\
                 \end{tabular} \\
                 \end{tabular}\\
\hline
CAP & 55 & \begin{tabular}{P{1.3cm} |P{2cm} }
                   910 & 910 \\
                 \end{tabular} \\
COVID-19 & 25 & \begin{tabular}{P{1.3cm}|P{2cm}}
                   3046 & 3046 \\
                 \end{tabular} \\
\hline
\end{tabular}
\label{tab1}
\end{table}


\subsection{Stage-1: Slice Label Prediction}
All slices for each patient in the dataset do not contain labels. Consequently, it is required to differentiate the slices with infections and without infections. To automatically detect the slices with infection and without infection, we use Stage-1 slice level label prediction framework (Fig. 2). We propose to independently label the unlabelled slices of COVID-19 and CAP 3D volumes using two different CNN architectures.   
We first used the SPGC dataset labels to create separate classes of infectious and non-infectious slices. Using transfer learning \cite{transfer}, we extracted features from the pre-trained CNN model for Stage-1. Several works have utilized transfer learning to extract features successfully for tasks both within and across domains  \cite{palak, sadbhawna, guntuku2015evaluating, guntuku2019twitter}. We utilized the weights of DenseNet-121  \cite{densenet} architecture to transfer learning across domains. The pre-trained features from Densenet-121 are obtained using Global Average Pooling (GAP) on the last layer. Then a fully connected layer with 1024 features is used. The fully connected layer was unfrozen, and its weights were trained. The extracted features were then fed into a Sigmoid activation function to obtain the two-class probabilities. 

We used Adam optimizer for the training process. The chosen parameters values for $\beta_1$ was 0.5 . The adopted learning rate for our model was $2*10^{-4}$. We obtained an accuracy of over 94 \% while fine-tuning. 

\subsection{Stage-2: Classification Model for Diagnosis}
After segmenting the 3D volumes into middle slices that exclusively contain lung data, as explained in the pre-processing step, we perform Stage-1 of the framework, slice-wise labeling of unlabelled slices. Our CNN architecture classifies CT scan slices into three fine-grained classes in the second stage: COVID-19, CAP, and healthy. The architecture designed for diagnosis purposes is illustrated in Fig. 3.

In order to classify CT scan images into three classes, we used several pre-trained architectures such as EfficientNet \cite{efficientnet}, InceptionV3 \cite{inception}, ResNet \cite{resnet}, and DenseNet \cite{densenet}. EfficientNet architecture has several advantages over other deep learning architectures such as compound scaling (in dimensions such as width, depth, resolution of an image, etc.), reduced set of parameters that make the training process efficient. For this reason, we fine-tuned EfficientNet architecture to extract features for our proposed classification model. These features are extracted using Global Average Pooling (GAP) and then fed into two fully connected dense layers with 2048 and 1024 trainable parameters, respectively. For three-class classification, a softmax activation function is used to predict the final class probabilities. The original and predicted labels (for COVID-19 and CAP) from Stage-1 are used for each of the three training classes. We make sure that there is no overlap in training and validation data and ensured out-of-sample patient validation. For training, we adopted the same optimizer as Stage-1 architecture i.e. Adam Optimizer with $\beta_1$ = 0.5 and learning rate of $2*10^{-4}$.

Single-slice prediction for final diagnosis does not guarantee the individual's diagnosis~\cite{covid-fact}. Hence, we utilized the slice-level predictions from Stage-2 of our proposed model for patient-level prediction of diagnosis based on a simple voting technique. All the slices of a patient are tested for classification into three classes. As discussed in Section 2.1, central slices of the 3D volume are important because they contain crucial information regarding infection and were used for patient-level prediction. Let $x$, $y$, $z$ be the total number of slices from the centre of the 3D volume predicted as COVID-19, CAP, normal, respectively. Further, let $x'$, $y'$, $z'$ be the total number of slices other than centre slices of 3D volume predicted as COVID-19, CAP, normal, respectively. We calculated the final Patient Label (PL) as:
$PL = (max ((x + 0.7x'), (y + 0.7y'), (z + 0.5z')))$,
where 0.7 and 0.5 are the weight factors obtained heuristically.

\section{RESULTS}

The proposed COVID+CAP-CNN model is trained and validated on different out-of-sample folds of the SPGC-COVID dataset. All analyses performed in Stage-2 are evaluated on non-overlapping out-of-sample sets. For an ablation study, we evaluated both the stages of our model for training and validation accuracy. Table 3 shows the accuracy of the proposed model for Stage-1, which is over 94 \%.

\begin{table}[h!]
\caption{Performance of Stage-1 for binary classification: infected and non-infected}
\label{table}
\centering
\small
\setlength{\tabcolsep}{3pt}
\begin{tabular}{|c||c|c|}
\hline
\textbf{Classification Type}& \textbf{Training accuracy} & \textbf{Validation accuracy} \\
\hline
COVID-19 & 99.29 \% & 97.7 \%\\
CAP & 99.17 \% & 94.7 \%\\
\hline
\end{tabular}
\label{tab1}
\end{table}

Stage-2 accuracy is shown in Tables 4 and 5. In Table-4, we show the variation of performance on different CNN architectures. EfficientNet is seen to outperform others. In Table-5, we evaluate the performance of the proposed model on different train-validation splits. Our proposed model achieved over 89.3 \%  accuracy on the validation set for different splits. 

We also evaluated patient-level accuracy for the three-way classification of the proposed framework and observed that the proposed algorithm achieves an 84\% accuracy across three fine-grained classes.

\begin{table}[t!]
\caption{Performance of Stage-2 for fine-grained classification: with different architectures.}
\label{table}
\centering
\small
\setlength{\tabcolsep}{2pt}
\begin{tabular}{|c||c|c|}
\hline
\textbf{Architecture}& \textbf{Training Accuracy} & \textbf{Validation Accuracy} \\
\hline
\textbf{Proposed(EfficientNet)}& \textbf{99.00 \%} & \textbf{89.3 \%}\\
InceptionV3 & 98.78 \% & 82.9 \% \\
ResNet & 99.61 \% & 78 \%\\
DenseNet & 98.66 \% & 80.6 \%\\
\hline
\end{tabular}
\label{tab1}
\end{table}

\begin{table}[t!]
\caption{Performance of Stage-2 for fine-grained classification: with different train-validation splits}
\label{table}
\centering
\small
\setlength{\tabcolsep}{2pt}
\begin{tabular}{|c||c|c|}
\hline
\textbf{Train-Validation Split}& 
\textbf{Training Accuracy} & \textbf{Validation Accuracy} \\
\hline
\textbf{70-30}& \textbf{99.00 \%} & \textbf{89.3 \%}\\
80-20 & 98.54 \% & 90.25 \% \\
90-10 & 99.3 \% & 91 \%\\
\hline
\end{tabular}
\label{tab1}
\end{table}

\subsection{Class-wise Sensitivity Analysis}

We show the confusion matrix considering the three classes: Normal, CAP, and COVID-19 in Fig. 4 and Fig. 5. The data provided in the confusion matrix in Fig. 4 is at slice-level, and that of Fig. 5 is at patient-level. Our fine-tuned EfficientNet model obtains a class-wise sensitivity of 89.2 \%, 81.9 \% and 91.49 \% for Normal, CAP and, COVID-19 classes, respectively. 

\begin{figure}[h!]
\centering
\subfloat{
  \includegraphics[width=.811\columnwidth]{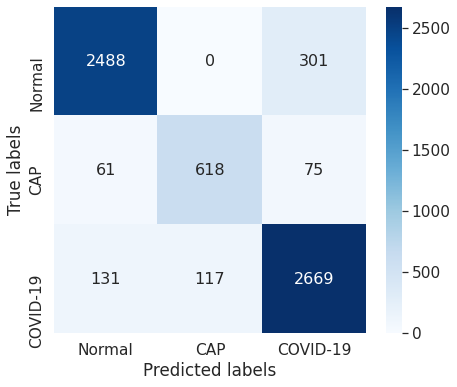}
}
\caption{Confusion Matrix for proposed proposed COVID+CAP-CNN model at slice-level.}
\end{figure}

\begin{figure}[h!]
\centering
\subfloat{
  \includegraphics[width=.811\columnwidth]{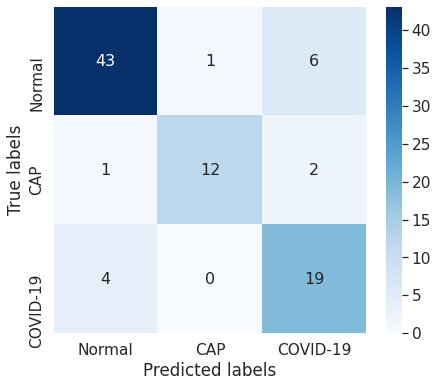}
}
\caption{Confusion Matrix for proposed proposed COVID+CAP-CNN model at patient-level.}
\end{figure}

\subsection{Execution Time}
We performed all the experiments on an NVIDIA Tesla V100 Tensor Core GPU. The proposed model takes less than 0.16 seconds for testing one CT scan image slice of shape $512\times512\times3$.

\section{CONCLUSION}
In this work, we proposed a two-stage framework to detect COVID-19 and CAP using CT scan images. In the first stage, individual slices of CT scans are labeled using fine-tuned DenseNet based deep-learning architecture. In the second stage, a fine-grained differential classification in three classes, i.e., COVID-19, CAP, and healthy individuals, by fine-tuning the EfficientNet architecture. The proposed two-stage framework achieved over 94\% accuracy for classifying the CT scan images in the binary classification task: infectious vs. non-infectious, and accuracy of 89.3\% for the fine-grained three-class classification: COVID-19, CAP, and normal. Code and model weights are available at ~\url{https://github.com/shubhamchaudhary2015/ct_covid19_cap_cnn}

\vfill\pagebreak
\newpage
 
 \bibliographystyle{IEEEbib}
\bibliography{ieeereferences}

\end{document}